# A Scoring Method for Driving Safety Credit Using Trajectory Data

Wenfu Wang
*College of Computer Science and Technology*
*Zhejiang University*
Hangzhou, China
wangwf73@zju.edu.cn

Weijie Yang
*College of Computer Science and Technology*
*Zhejiang University*
Hangzhou, China
yangwj92@ zju.edu.cn

An Chen
*College of Computer Science and Technology*
*Zhejiang University*
Hangzhou, China
anchencn@126.com

Zhijie Pan
*College of Computer Science and Technology*
*Zhejiang University*
Hangzhou, China
zhijie_pan@zju.edu.cn

*Abstract*—Urban traffic systems worldwide are suffering from severe traffic safety problems. Traffic safety is affected by many complex factors, and heavily related to all drivers' behaviors involved in traffic system. Drivers with aggressive driving behaviors increase the risk of traffic accidents. In order to manage the safety level of traffic system, we propose *Driving Safety Credit* inspired by credit score in financial security field, and design a scoring method using driver's trajectory data and violation records. First, we extract driving habits, aggressive driving behaviors and traffic violation behaviors from driver's trajectories and traffic violation records. Next, we train a classification model to filtered out irrelevant features. And at last, we score each driver with selected features. We verify our proposed scoring method using 40 days of traffic simulation, and proves the effectiveness of our scoring method.

*Keywords—driving safety credit, credit scoring, random forest, trajectory data mining, driving behavior analysis*

## I. Introduction

Urban traffic worldwide is facing severe traffic safety problems. According to the *Global status report on road safety 2015* [1] released by the World Health Organization, approximately 1.3 million people die each year on the world's roads, and between 20 and 50 million sustain non-fatal injuries.

Traffic system is one example of highly complex systems, and its influencing factors are numerous. For traffic safety, drivers' behaviors are the primary cause, especially aggressive driving behaviors. Evans [2] found that driver-related behavioral factors dominate the causation of a motor vehicle accident and contribute to the occurrence of 95% of all accidents. Petridou's study [3] tried to further clssified driving behavioral factors associated with accidents. In particular, aggressive driving behaviors has strong correlation with traffic accidents, and has been perceived as one of the most serious problems in modern day driving [4]. [5] estimated that aggressive driving was the main cause of majority of accidents from 2003 to 2007 in United States.

NHTSA defines aggressive driving as "directly affects other road users by placing them in unnecessary danger" [6], similar definition also in [7]. Aggressive driving may include excessive speeding, tailgating, abrupt lane changing [7]. Many risky driving behaviors are habitual and thus are especially dangerous, such as habitual speeding, habitual disregard of traffic regulations, risk taking behaviors and so on. As Williams [8] showed in his study that highly skilled drivers in the USA had much higher accident rate, because they were confident about their skills and often took risky driving behaviors. These researches show that driving behaviors have significant impact on driving performanece and therefore on the possibility of an accident. And it could be expected that safe driving behaviors from each driver are of great help for improving traffic safety.

It is crucial to provide constructive feedback to drivers to help them correct unsafe driving behaviors [9]. Traditionally, demerit point system are widely adopted. Enforcement endorse heavy penalty points to unsafe driving behaviors. However, increasingly strict regulations and heavy punishment are not as effective as it would be. To promote safe driving behaviors, we propose *Driving Safety Credit* to quantitatively evaluate each driver's safety level. *Driving Safety Credit* is inspired by credit score in financial field. Banks use credit score to assess the potential risk posed by lending money to borrowers and to manage bad debt risk. In the field of credit scoring, the objective is to build models that can extract knowledge of credit risk evaluation from past observations and to apply it to evaluate credit risk in future [10]. Similarly, we propose a scoring method to score *Driving Safety Credit* using a driver's past driving histories and apply it to evaluate the future driving behavior risk. We believe driving habits, aggressive driving behaviors and traffic violation records could fully cover the driving history of a driver. Our scoring method consists of two stages: feature extraction and credit scoring. In feature extraction stage, we use driving habits, aggressive driving behaviors and traffic violation records as features which are risk taking behaviors in driving [3] , and we extract them from trajectory data of each driver. In credit scoring stage, we first train a classification model to classify drivers into good or bad, then filter out small weight features, and then we derive the final *Driving Safety Credit* based on selected features.

With the popularity of smartphones, upcoming connected cars and development of large scale machine learning methods, *Driving Safety Credit* is not only feasible but also potentially important in building smart transportation systems and smart

Supported by National Natural Science Foundation of China (Grant No. 61751209)

cities. *Driving Safety Credit* derives from one's driving histories, and will be much more effective than demerit point system. *Driving Safety Credit* may also be integrated into personal credit system, and governments could leverage it to assign different driving rights to drivers, thus promoting drivers towards safe driving behaviors.

The remainder of this paper is organized as follows: Section II provides a literature review on driving behaviors analysis and credit scoring. Section III first introduces the framework of our scoring method, and then details feature extraction from trajectory data and credit scoring using random forest. Section IV reports simulation experiment setup and discuss the model selection, imbalance dataset, and scoring results. Finally, section V draws our conclusions and discusses the future work.

## II. RELATED WORK

Our scoring method lies an intersection between driving behavior analysis and credit scoring. We briefly introduce the two research fields.

### A. Driving behaviors analysis

Smartphones and IOT devices are prevalent in people's life, with rich data generated by various sensors, they are becoming the perfect proxy to quantitatively study human driving behaviors. There are abundant researches analyzing driving behaviors using data collected from smartphone sensors or GPS. In general, they extract features, such as acceleration, deceleration and turning, handcraft templates or penalty points to classify driving behavior or score a trip [9], [11]–[15]. In [15], Hong used samrtphone and in-vehicle data to analyze aggressive driving style. Fazeen [9] used the three-axis accelerometer on an Android smartphone to record the data, and used the data to analyze driver behaviors. Unsafe acceleration, decleration and sudden lane changing driving patterns are analyzed. Eren [11] obtained position, speed, acceleration, deceleration, and deflection angle sensory information from accelerometer, gyroscope and magnetometer, and then used Bayesian classification algorithm to classify safe/unsafe driving style. Specially, in order to accurately detect driving events, such as abrupt acceleration, abrupt deceleration and unsafe turn, they manually designed templates to match driving events. In [12], Castignani designed a Fuzzy Inference System to detect risky events using data from accelerometer, magnetometer, gravity sensor and GPS, and simply substructed points according to the risky events during the trip. The fuzzy rules and penalty points were manually designed, and the score was on a driver's one trip.

Driving behaviors analysis needs large scale dataset and machine learning methods. The above researches all conducted in small scale and focused on one trip. [11] had 15 drivers for his experiments, [14] had 72 valid participants, [12] used a single driver and four different runs for calibration and ten drivers for scoring. IMU sensors data are also biased due to its limited precision. Paefgen [14] found that mobile measurements from IMU tend to overestimate critical driving events, and he claimed that it was possibly due to deviation from the calibrated initial device pose.

Trajectory data mining promises to quantitatively analyze traffic system in big data era [16]. In [17], Wang collected the GPS trajectory of more than 32,000 taxis over a period of two months, and used it to predict travel time on a trip. Rich driving information, such as acceleration and deceleration can also be extracted through trajectory data. Ren [18] derived sharp acceleration, deceleration and abrupt turns from GPS trajectory data, and analyzed driving behavior of bus drivers. We adpot trajectory data as the main part of our dataset and extract driving behavioral features from it.

We believe large scale trajectory dataset collected in real world with diverse driving styles is vital to assess drivers' driving behaviors. However, we find that such dataset does not exist. [17] collected a large trajectory dataset, but only from taxi drivers. We instead turn to traffic simulation to generate our dataset. We carefully design the driving styles in our simulation. This paper focus more on the scoring method of *Driving Safety Credit*, and we put collecting real world data and test the effectiveness *Driving Safety Credit* of into our future work.

### B. Credit scoring

In field of credit scoring, there are observation period and performance period. The data obtained during observation period are used to extract feature, and data obtained during performance period are used to construct label [19]. Then the scoring model uses the feature to predict its label. Since 1990s, machine learning techniques have been studied extensive as tools for credit score modeling, such as logistic regression (LR), support vector machine (SVM), decision tree (DT), random forest (RF), neural network (NN) [20]. There are two problems that needs more research attention, one is feature selection and the other is imbalanced dataset.

Feature selection is a critical task in credit scoring, good features have considerable impact on improving prediction performance in credit scoring. However, there is no consensus upon the most representative feature. Usually, variables are first examined in order to evaluate their importance and explanatory power in the dataset. Many studies didn't apply feature selection, and directly employed original features provided in the datasets or selected the features according to the advice of domain experts [20]. In general, performing the genetic algorithm and logistic regression for feature selection can provide prediction improvement [21].

Another problem in credit scoring field is imbalanced dataset. If the ratio of negative to positive samples evidently deviated from the actual proportion, the model's prediction capability would be distorted. Currently, popular solutions for imbalanced datasets include over-sampling, under-sampling and a hybrid of them at data level, and adjusts the predictive modelling by introducing a threshold or cost-sensitive learning at model level. With imbalanced datasets, assessing and predictive models may provide misleading information. For example, a model can provide 99% accuracy when predicting all the samples as positive, if the dataset consists of 99% positive samples and 1% negative samples. Hence, researches based on imbalanced datasets should pay more attention about Type I and II error rates [22].

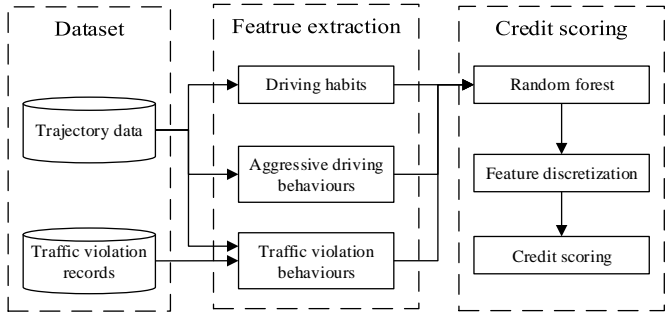

Fig. 1. The framework of our credit scoring method

III. FRAMEWORK

*A. Framework*

Fig. 1 describes the framework our work. It consists of two stages: feature extraction and credit scoring.

In feature extraction stage, features are extracted from our own dataset, including trajectory data and traffic violation records. Features are divided into three categories: driving habits, aggressive driving behaviors and traffic violation behaviors. Driving habits mainly describe the driver's driving characteristic under normal circumstances, such as average acceleration, average deceleration, max acceleration, and max deceleration . Aggressive driving behaviors include abrupt acceleration, abrupt deceleration and abrupt turning and so on. Traffic violation behaviors are the impassable baseline of traffic safety, a driver is not supposed to violate traffic regulation. If a driver has seriously violated traffic rules, he or she is considered risky to traffic system safety. Traffic violation includes speeding, traffic light violation and collision, etc. We believe driving habits, aggressive driving behaviors and traffic violation records have effectively covered a driver's driving history.

In creadit scoring stage, there are two tasks: classification task and scoring task. In the classification task, three categories driving behaviors in observation period are used as features. And if a driver have traffic violation behaviors in performance period, he or she is labelled bad drivers, otherwize good drivers. Pairing features and labels, a classification model is trained to classify drivers into good or bad. Then the weight for every feature can be got from the model. In the scoring task, the score for each interval of every feature is calculated. And finally, *Driving Safety Credit* is derived above selected features and weights.

We list the measurements abbreviations used throught this paper, shown in TABLE I. We also define trajectory as follows, which adopts trajectory definition in [16]:

**Definition** (*Trajectory*): A is a trace generated by a moving object in geographical spaces, denoted by $T_r$. Usually, it is represented by a series of chronologically ordered points, for example, $p_1 \rightarrow p_2 \rightarrow \cdots p_n$. And each point $p$ contains data as follow: time *t*, speed *v*, longitude *lng*, latitude *lat*, angle or orientation *h* and driver's unique id *u*, e.g. $p = \{t, v, lng, lat, h, u\}$.

TABLE I. Measurements abbreviations

| Abbreviations | Description |
|---|---|
| $a_k$ | $a_k$ is the acceleration at point $p_k$ (m/s$^2$) |
| AVGT | Average driving time of each trip |
| AVGS | Average driving distance of each trip |
| MAXA | Maximum acceleration of a driver |
| AVGA | Average acceleration of a driver |
| MAXD | Maximum deceleration of a driver |
| AVGD | Average deceleration of a driver |
| ISN | Total number of crossing intersection |
| ACC | Abrupt acceleration threshold |
| DEC | Abrupt deceleration threshold |
| V*, ANG | When speed greater than V* and the turning angle greater than ANG, abrupt turning happens |
| AAS | Total driving distance of abrupt acceleration |
| AAT | Total driving time of abrupt acceleration |
| AAN | The number of abrupt acceleration |
| ADS | Total driving distance of abrupt deceleration |
| ADT | Total driving time of abrupt deceleration |
| AND | The number of abrupt deceleration |
| ATS | Total driving distance of abrupt turning |
| ATT | Total driving time of abrupt turning |
| ATN | The total number of abrupt turning |
| OSS | Total driving distance of over speed |
| OST | Total driving time of over speed |
| OSN | The number of over speed |
| $W_i$ | The weight of feature *i* |
| $NW_i$ | The normalized weight of feature *i* |

*B. Feature Extraction Stage*

In this paper, nine representative features can be extracted from trajectory data, including *AVGT, AVGS, MAXA, AVGA, MAXD, AVGD, MAXV, AVGV,* and *ISN*. *AVGT* is calculated as the average value of trips' driving time. And the *AVGS* is calculated as the average value of trips' driving distance. $a_k$ is calculated using following equation:

$$a_k = (p_k \cdot v - p_{k-1} \cdot v)/(p_k \cdot t - p_{k-1} \cdot t) \qquad (1)$$

When data sampling frequency is low, there will be great error between the calculated acceleration and the real acceleration. As a result, it is necessary to use data with high frequency. In this work, the data sampling frequency is 1 Hz. Although deceleration is calculated using the same method as acceleration, it reflects different driving behaviours.

Three kinds of aggressive driving behaviours are extracted from trajectory data, including abrupt acceleration, abrupt deceleration and abrupt turning. When the acceleration value is greater than threshold *ACC*, it is considered as abrupt acceleration. Similarly, when the deceleration value is greater than threshold *DEC*, it is considered as abrupt deceleration. And when the turning angle greater than *ANG* as well as the speed is over threshold *V\**, it is considered as abrupt turning. Same as [18], we use total driving distance, total driving time and the total number to describe abrupt acceleration, abrupt deceleration, abrupt turning.

If trajectory $T_r$: $\{p_1, p_2, ..., p_n\}$ contains abrupt acceleration, then,

$$AAS = AAS + \text{distance}(p_1, p_n) \quad (2)$$

$$AAT = AAT + p_n \cdot t - p_1 \cdot t \quad (3)$$

$$AAN = ANN + 1 \quad (4)$$

If trajectory $T_r$: $\{p_1, p_2, ..., p_n\}$ contains abrupt deceleration, then,

$$ADS = ADS + \text{distance}(p_1, p_n) \quad (5)$$

$$ADT = ADT + p_n \cdot t - p_1 \cdot t \quad (6)$$

$$ADN = ADN + 1 \quad (7)$$

If trajectory $T_r$: $\{p_1, p_2, ..., p_n\}$ contains abrupt turning, then,

$$ATS = ATS + \text{distance}(p_1, p_n) \quad (8)$$

$$ATT = ATT + p_n \cdot t - p_1 \cdot t \quad (9)$$

$$ATN = ATN + 1 \quad (10)$$

In this paper, speeding, traffic light violation and collision are recorded as traffic violation behaviours due to the simulation limitation. Total times is used to measure traffic light violation and collision. Total driving time *OST*, total driving distance *OSS*, total times *OSN* are used to measure speeding.

If trajectory $T_r$: $\{p_1, p_2, ..., p_n\}$ contains speeding, then,

$$OSS = OSS + \text{distance}(p_1, p_n) \quad (8)$$

$$OST = OST + p_n \cdot t - p_1 \cdot t \quad (9)$$

$$OSN = OSN + 1 \quad (10)$$

*C. Credit Scoring Stage*

In credit scoring stage, we first train a classification model to determine the weight of different features. Generally, the greater weight shows that the feature is more important. Then, we filter out features with small weight. Finally, we can score drivers based on the selected features.

The main objective of classification model is to classify drivers into good or bad. Classification is a fundamental topic in machine learning research field, and many methods are developed to solve it. In this paper, we choose RF as our classification model. We will conduct experiments to compare different classification models, and RF outperforms other models.

RF is a combination of tree predictors such that each tree depends on the values of a random vector sampled independently and with the same distribution for all trees in the forest. More details about RF please refer to [23]. What's most important about RF is that it is robust to noise, which means RF can still be accurate even with feature losses.

Before training RF model, data should be resampled to keep the ratio between the positive and negative class balanced. We also utilize downsampling. After training, the RF model will output the weight of each feature. If a feature get greater weight, it is regarded more significant The weight set is represented as $\{W_i\}$, where $i$ belongs to *S*, and *S* is the feature set. Features with small weight are filtered out. Finally, for a feature $i$ in reserved set $S^*$, its new weight $NW_i$ is normalized as follows:

$$NW_i = \frac{W_i}{sum_{k \in S^*}(W_k)} \cdot 100 \quad (14)$$

Before scoring drivers, in order to differentiate the value of feature properly, value range of each feature is divided into three intervals and each interval is assigned a score value. In particular, continuous variables need to be discretized. When discretizing, it is necessary to keep the entropy as small as possible.

To assign values to each interval, the proportion of bad driver $p_{i,j}$ should be calculated first. Then the score assignment to each interval is determined as follows:

$$f_{i,j} = \frac{1 - p_{i,j}}{\max_{k=1}^{n_i}(1 - p_{i,k})} \quad (15)$$

For feature $i$, when the value falls into interval $j$, its score is:

$$h_{i,j} = f_{i,j} \cdot NW_i \quad (16)$$

Finally, the score of driver $u$, i.e. *Driving Safety Credit*, is derived using the following equation:

$$score_u = \sum_{i \in S^*} h_{i,j} \quad (17)$$

To address the imbalance data problem, we perform two methods to our dataset. First, the dataset is resampled and tested under different ratios. Second, the precision of the model is defined as the proportion of labelled good drivers in the predicted good drivers, which is used to evaluate the classification model.

IV. SIMULATION & ANALYSIS

We simulate the traffic of Hangzhou, China with 22631 drivers over 40 days. The trajectory data of simulation consist of our dataset. The simulation setup will be described in detail. We conduct expriments to show RF model achieves best balance of accuracy and robutness. We also conduct experiments to demonstrate the effect of imbalance dataset. And finally, we prove our scoring method.

*A. Simulation Setup*

Simulation of Urban Mobility (SUMO) software is chosen as our simulation tool. SUMO is an open-source, microscopic road traffic simulation software [24]. It can simulate various types of transportation vehicles with different driving styles. In SUMO, vehicles are independent, which means that every vehicle has their own route and moves individually through the road network. It is a space continuous, time discrete (the default duration of each time step is one second) system. Although traffic light violation is not supported in SUMO, the deceleration of vehicle at a traffic light recorded by SUMO is used as a proxy to obtain traffic light violation. When the deceleration of a vehicle at a traffic light is larger than the threshold, it is regarded as traffic light violation behaviours.

TABLE II. The value of standard parameters in SUMO

| acc (m/s$^2$) | dec (m/s$^2$) | $S_{max}$ (m/s) | $G_{min}$ (m) | $\tau$ (s) |
|---|---|---|---|---|
| 2.6 | 4.5 | 70 | 2.5 | 1.0 |

TABLE III. Driver styles designed in this paper

| No | acc (m/s$^2$) | dec (m/s$^2$) | $\sigma$ | $S_{max}$ (m/s) | $G_{min}$ (m) | $\tau$ (s) | PR |
|---|---|---|---|---|---|---|---|
| 1 | 2.5 | 2.0 | 0.5 | 23 | 2.6 | 1.2 | 8% |
| 2 | 2.4 | 2.5 | 0.5 | 23 | 2.7 | 1.3 | 10% |
| 3 | 3.1 | 3.5 | 0.6 | 33 | 1.2 | 1.0 | 12% |
| 4 | 3.0 | 3.4 | 0.6 | 33 | 1.3 | 1.0 | 10% |
| 5 | 2.8 | 2.6 | 0.55 | 21 | 2.8 | 1.5 | 12% |
| 6 | 2.6 | 2.5 | 0.55 | 21 | 2.9 | 1.7 | 14% |
| 7 | 2.9 | 3.6 | 0.64 | 28 | 1.5 | 1.2 | 8% |
| 8 | 2.7 | 3.4 | 0.62 | 28 | 1.6 | 1.3 | 6% |
| 9 | 2.3 | 2.8 | 0.53 | 19 | 2.6 | 1.9 | 8% |
| 10 | 2.2 | 2.9 | 0.52 | 19 | 2.8 | 2.0 | 9% |
| 11 | 2.6 | 3.3 | 0.59 | 25 | 1.8 | 1.3 | 2% |
| 12 | 2.4 | 3.1 | 0.58 | 25 | 2 | 1.4 | 1% |

TABLE IV. Gauss noises (mean and standard derivation)

| acc (m/s$^2$) | dec (m/s$^2$) | $\sigma$ | $S_{max}$ (m/s) | $G_{min}$ (m) | $\tau$ |
|---|---|---|---|---|---|
| (0, 0.15) | (0, 0.15) | (0, 0.01) | (2, 1) | (0, 0.1) | (0.2, 0.05) |

The value of standard SUMO simulation parameters related to driver styles are listed in TABLE II. The abbreviations of terms in TABLE II are: *acc* for acceleration, *dec* for deceleration, $G_{min}$ for the minimum gap acceptance, $S_{max}$ for maximum speed and $\tau$ for a driver's reaction time. If $\tau$ value is smaller than 1, unrealistic results will be generated [25]. So $\tau$ is always greater than 1 in our simulation.

Referring to [25], 12 driver styles are designed in our simulation, the designed driver styles are listed in TABLE III. $\sigma$ means the driver imperfection value, its value is between 0 and 1. And *PR* in TABLE III means the proportion for a driver belongs to this driver style. Note that we should pay more attention to the differences of parameters of driver styles rather than the parameters themselves.

With above driver styles, 22631 drivers are randomly generated in total. Each driver has his own route and each trip is no less than 3km. To make the driving style more stochastic, Gauss noises were added to the parameter values, the mean and standard derivation for each parameter are described TABLE IV. It should be noted that the value of $\tau$ must not be less than 1. 40 days of traffic are simulated, and each day is from 6:00 to 10:00. The first 20 days are treated as the observation period, and the second 20 days are treated as the performance period. Trajectory points are sampled by 1 Hz. We totally get 21305 "good" drivers and 1326 "bad" drivers.

Given the trajectory points, we further processed to derive nine representative features: *AVGT, AVGS, MAXA, AVGA, MAXD, AVGD, MAXV, AVGV,* and *ISN*. The equations used to derive features are described in Section III.

### B. Classification Model Selection

We compare the classification resulst of random forest (RF), logistic regression (LR), decision tree (DT), Naïve Bayes Model (NB). We use K-fold cross-validation to validate the performance of these models. In the experiment, the ratio of positive and negative samples was set to 1:1. Accuracy is defined as the ratio between the number of correct classified samples and the number of test samples. A model is better if it gets higher accuracy.

Fig. 2 shows the accuracy of different models. As we can see, RF always gives a better accuracy than LR and DT.

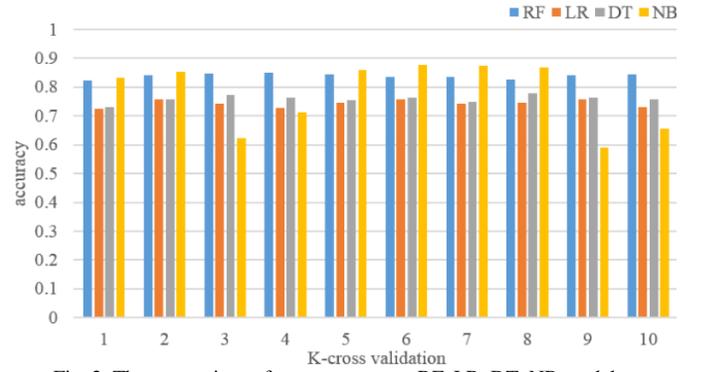
Fig. 2. The comparison of accuracy among RF, LR, DT, NB models

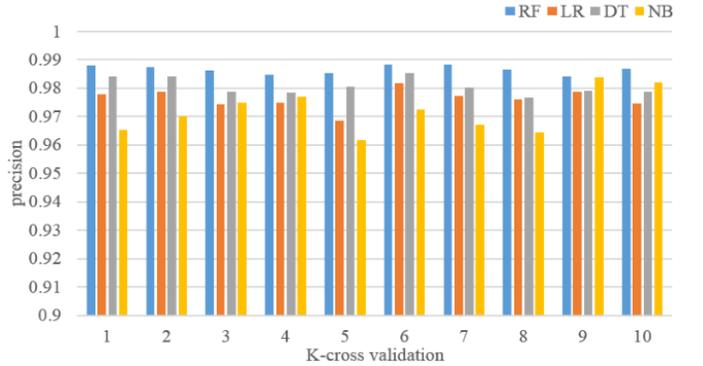
Fig. 3. The comparison of precision of good drivers among RF, LR, DT, NB models

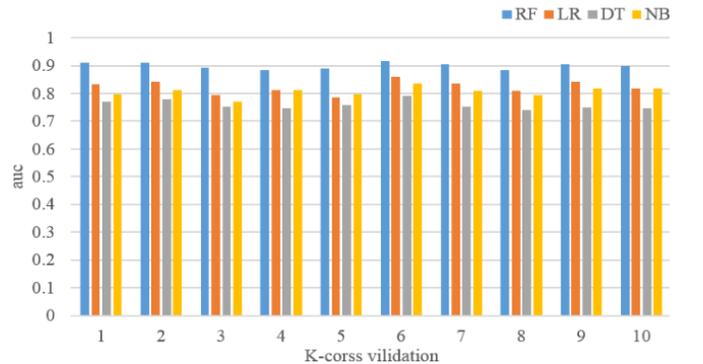
Fig. 4. The comparison of AUC among RF, LR, DT, NB models

In some cases, NB gives a better result than random forest while sometimes NB is the worst model. Among the four models, RF achieves the best balance of accuracy and robustness.

In the credit scoring field, it's worse to classify the bad consumer to good than classify the good consumer to bad. And it is the same for traffic safety credit. It's worse to classify the bad driver as good driver than classify the good driver as bad driver. Taking that into consideration, we take the precision of good driver as the criteria of the model, which is the ratio of the number of labeled good drivers and the number of predicted good drivers. A model has better performance if it has higher precision. Fig. 3 shows that the precision of good drivers for different models. As we can see, RF achieves highest precision all the time which also proves its robustness.

In machine learning, area under curve (AUC) is another important measurement. A higher AUC value means a model is more robust. We calculate the AUC value of different models, and the result is shown in Fig. 4. RF again outperforms other models. Therefore, we choose RF as our classification models.

### C. Imbalanced Dataset

We also conduct experiments to show the effect of different ratio of positive and negtive smaples. In our dataset, we get 21305 "good" drivers and 1326 "bad" drivers. The ratio of positive and negative samples is about 16:1, this imbalanced ratio will heavily infulence our RF model performance, and it is important to deal with this imbalanced dataset problem. We conduct expriements to show the impact of imbalanced dataset.

We resample the dataset to obtain 1:10, 1:2, 1:1, 2:1, 4:1, 8:1, and 10:1 positive to negtive ratio, and test our RF model. The results under different ratios are shown in Fig. 5. We can see that with the increase of positive samples, the accuracy increases. However, this high accuracy is misleading. In Fig. 6, we use precision to measure the effect of different ratios. We can see that with the increase of ratio, the precision decreases. The main reason is that with the increase of positive samples, a test sample is more likely to be classified as good than bad. Therefore, in order to achieve balance between accuracy and precision, we set our positive to negtive ratio as 1:1.

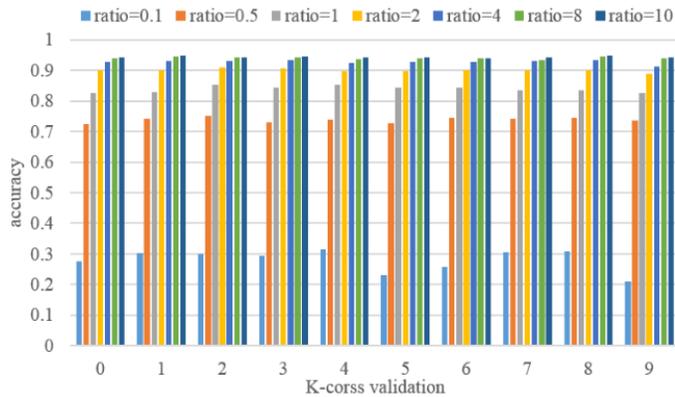

Fig. 5. Accuracy of RF model under different ratios

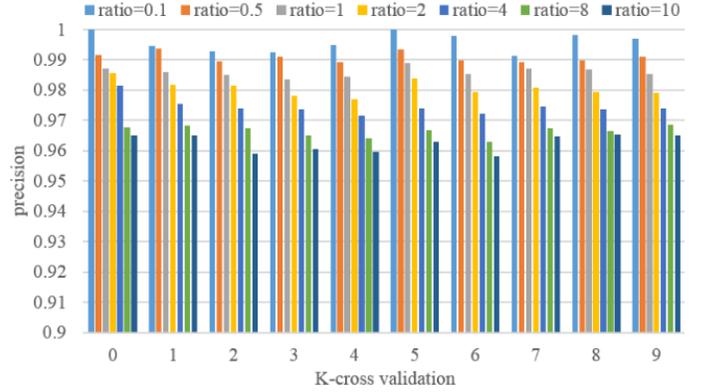

Fig. 6. Precision of RF model under different ratios

### D. Driving Safety Credit Scoring

We use the proportion of bad drivers in the top N drivers as metric to measure. With a fixed N, the scoring model is better if proportion of bad drivers is lower. While with a fixed proportion, the scoring method is better if N is smaller. The result is shown in TABLE V. In the top 1000, only 1 driver is bad. While in the top 5000, 14 drivers are bad, and in the top 10000, 56 drivers are bad.

A intuitive illustration of TABLE V is shown in Fig. 7. The main axis represents the range of *Driving Safety Credit* score interval, and the secondary axis represents the range of *Driving Safety Credit* rank interval. The interval lies in the middle represents the number of bad drivers in corresponding range. As depicted in Fig. 7, the majority of bad drivers, that is 88.61%, are scored in [15000, 22631]. This result shows that our proposed scoring method has good performance.

TABLE V. The credit score range and the number of bad drivers in each interval of credit rank

| Credit rank interval | Credit score range | The number of bad drivers | Proportion of total bad drivers |
|---|---|---|---|
| [1,500) | [100,82.29) | 0 | 0 |
| [500,1000) | [82.29,80.80) | 1 | 0.075% |
| [1000,5000) | [80.80,76.61) | 13 | 0.98% |
| [5000,10000) | [76.61,68.55) | 42 | 1.81% |
| [10000,15000) | [68.55,64.51) | 95 | 7.16% |
| [15000,20000) | [64.51,51.41) | 444 | 33.48% |
| [20000,22631) | [51.41,0] | 731 | 55.13% |

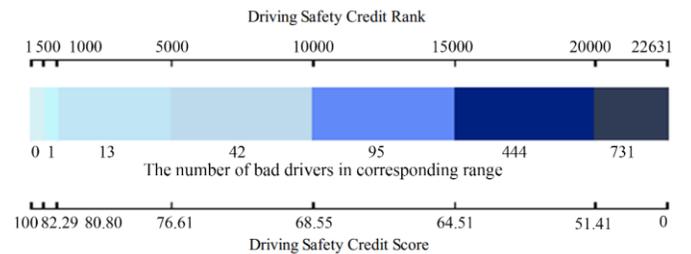

Fig. 7. Illustration of TABLE V

## V. CONCLUSIONS

Driving behavioral factors are the dominant factor of traffic accidents. To promote safe driving behaviors, we propose *Driving Safety Credit* to quantitatively evaluate each driver's safety level. We generate a large scale trajectory dataset with 22631 drivers over 40 days using SUMO simulation software, and extract driving habits, aggressive driving behaviours and traffic violation behaviors from it. Given features, we use our scoring method to obtain *Driving Safety Credit*. The result that 88.61% bad drivers rank in [15000, 22631] range proves our proposed method is effective for scoring *Driving Safety Credit*.

In the future, we plan to take more personal features into consideration to score *Driving Safety Credit* more comprehensively, such as age, gender, and education etc. Furthermore, we also would like to collect large scale real world data to further test our proposed methods.

ACKNOWLEDGMENT

I would like to thank Professor Pan for his advice in the discussion of this paper, and An Chen and Weijie Yang for their work on this paper.